\documentclass[11pt,a4paper,preprint]{article}
\pdfoutput=1
\usepackage{jheppub} 
\usepackage{graphicx}
\usepackage{graphics}
\usepackage[mathscr]{eucal}
\usepackage{amssymb}
\usepackage{amsmath}
\usepackage{xspace}
\usepackage{listings}
\usepackage{ulem}
\usepackage{hyperref}
\usepackage[capitalise, english]{cleveref}
\usepackage{multirow}
\usepackage{booktabs}
\usepackage[exponent-product = \times]{siunitx}
\usepackage{bbold}
\usepackage{soul}
\usepackage{color}
\usepackage{centernot}
\usepackage{enumerate}
\usepackage{makecell}

\def\da{\dagger}
\def\pa{\partial}

\def\beqa{\begin{eqnarray}}
\def\eeqa{\end{eqnarray}}

\def\nn{\nonumber}

\def\[{\left[}
\def\]{\right]}
\def\({\left(}
\def\){\right)}

\def\..{\left.}
\def\.{\right.}

\def\ra{\rightarrow}

\def\tm{\times}

\newcommand{\BDi}{\ensuremath{B_D^{(i,i)}}}

\title{\sffamily  Scalar dark matter explanation of the DAMPE data in the minimal Left-Right symmetric model}
\author[a,b]{Junjie Cao}
\author[a]{,Xiaofei Guo}
\author[a]{,Liangliang Shang}
\author[a,c]{,Fei Wang}
\author[d]{,Peiwen Wu}
\author[e,f]{,Lei Zu}
\affiliation[a]{College of Physics and Materials Science, Henan Normal University, Xinxiang 453007, China}
\affiliation[b]{Center for High Energy Physics, Peking University, Beijing 100871, China}
\affiliation[c]{School of Physics, Zhengzhou University, 450000, ZhengZhou, P.R.China}
\affiliation[d]{School of Physics, KIAS, 85 Hoegiro, Seoul 02455, Republic of  Korea}
\affiliation[e]{Key Laboratory of Dark Matter and Space Astronomy, Purple Mountain Observatory, Chinese Academy of Sciences, Nanjing 210008, China}
\affiliation[f]{School of Astronomy and Space Science, University of Science and Technology of China, Hefei 230026, Anhui, China}
\emailAdd{junjiec@itp.ac.cn}
\emailAdd{guoxf@gs.zzu.edu.cn}
\emailAdd{shlwell1988@foxmail.com}
\emailAdd{feiwang@zzu.edu.cn}
\emailAdd{pwwu@kias.re.kr}
\emailAdd{zulei@pmo.ac.cn}

\abstract{Left-Right symmetric model (LRSM) has been an attractive extension of the Standard Model (SM) which can address the origin of parity violation in the SM electroweak (EW) interactions, generate tiny neutrino masses, accommodate dark matter (DM) candidates and provide a natural framework for baryogenesis through leptogenesis. In this work we utilize the minimal LRSM to study the recently reported DAMPE results of cosmic $e^+e^-$ spectrum which exhibits a tentative peak around 1.4 TeV, while satisfying the current neutrino data.
We propose to explain the DAMPE peak with a complex scalar DM $\chi$ in two scenarios: 1) $\chi\chi^* \to H_1^{++}H_1^{--} \to \ell_i^+\ell_i^+\ell_j^-\ell_j^-$; 2) $\chi\chi^* \to H_{k}^{++}H_{k}^{--} \to \ell_i^+\ell_i^+\ell_j^-\ell_j^-$ accompanied by $\chi\chi^* \to H_1^+ H_1^- \to \ell_i^+ \nu_{\ell_i} \ell_j^- \nu_{\ell_j}$ with $\ell_{i,j}=e,\mu,\tau$ and $k=1,2$. We
fit the theoretical prediction on $e^+e^-$ spectrum to relevant experimental data to determine the scalar mass spectrum favored by the DAMPE excess. We also consider
various constraints from theoretical principles, collider experiments as well as DM relic density and direct search experiments.
We find that there are ample parameter space which can interpret the DAMPE data while passing the constraints. Our explanations, on the other hand, usually
imply the existence of other new physics at the energy scale ranging from $10^7 {\rm GeV}$ to $10^{11} {\rm GeV}$.  Collider tests of our explanations are also discussed.}

\begin{document}
\maketitle \indent
\newpage

\section{\label{introduction}Introduction}
The discovery of the Higgs boson at the Large Hadron Collider (LHC) indicates that the Standard Model (SM) of particle physics is a highly successful theory in describing a large amount of low energy phenomena \cite{Aad:2012tfa,Chatrchyan:2012xdj}. On the other hand, the origin of the chiral structure of the SM, which is crucial in understanding why the SM matter contents are much lighter than the Planck scale, is not explained in the framework of SM. In fact, it is still unknown so far why the weak interaction violates parity while all other interactions conserve parity and whether parity conservation can be achieved at a more fundamental level.

Left-Right symmetric model (LRSM), which is a vector extension of the SM with an enlarged gauge group $SU(3)_C \times SU(2)_L \times SU(2)_R \times U(1)_{B-L}$ \cite{Pati:1974yy,Mohapatra:1974hk,Mohapatra:1974gc,Senjanovic:1975rk,Senjanovic:1978ev}, assumes that the fundamental weak interaction is invariant under parity symmetry and the observed parity violation is the consequence of the spontaneous breaking of parity symmetry. Requiring the existence of right-handed neutrinos in the LRSM, tiny neutrino masses can naturally be generated by Type II seesaw mechanism \cite{Minkowski:1977sc,Mohapatra:1979ia,Lazarides:1980nt,Mohapatra:1980yp,Schechter:1980gr}. Besides, the LRSM can accommodate dark matter candidates \cite{Berlin:2016eem,Borah:2016uoi,Berlin:2016hqw,Patra:2015vmp,Garcia-Cely:2015quu,Heeck:2015qra,Bandyopadhyay:2017uwc,Dev:2016xcp,Dev:2016qeb} and provide a natural framework for baryogenesis through leptogenesis \cite{Fukugita:1986hr}. Its gauge group can naturally appear in typical $SO(10)$ GUT group breaking chain $SO(10)\ra SU(4)_{PS}\tm SU(2)_L\tm SU(2)_R\ra LR$ or from the breaking of some other partial unification theories such as $SU(4)_{PS} \tm SU(4)_W,SU(7)$ etc \cite{Li:2009sw,Balazs:2009kj,Balazs:2010yw}. The $SU(2)_R\tm U(1)_{B-L}$ breaking scale, which can be characterized by the $W_R$ gauge boson masses, is well motivated to lie as low as several TeV \cite{Chang:1983fu,Chang:1984uy,Arbelaez:2013nga,Bambhaniya:2014cia,Maiezza:2016bzp,Maiezza:2016ybz,Chakrabortty:2016wkl} and provides us the search possibilities in various physical experiments, such as collider signals \cite{Bambhaniya:2015wna,Bambhaniya:2013wza,Keung:1983uu,Chiappetta:1993jy,Maiezza:2010ic,Tello:2010am,Nemevsek:2011hz,Esteves:2011gk,Das:2012ii,Chen:2013fna,Arbelaez:2013nga,Das:2016akd,Das:2017hmg,Dev:2016dja,Dev:2016vle,Dev:2017dui}, flavor observables \cite{Gluza:2015goa,Gluza:2016qqv,Deshpande:1990ip,Hirsch:1996qw,Awasthi:2013ff,Barry:2013xxa,Awasthi:2015ota,Bambhaniya:2015ipg,Tello:2010am,Zhang:2007da,Guadagnoli:2010sd,Blanke:2011ry,Bertolini:2014sua}, as well as EW precision parameters \cite{Czakon:1999ue,Chakrabortty:2012pp}.

Very recently, the DArk Matter Particle Explorer (DAMPE) experiment reported new results of the total cosmic $e^++e^-$ flux measurement between 25 GeV and 4.6 TeV,
which contain a spectral softening at around 0.9 TeV and a peak at around 1.4 TeV \cite{Collaboration2017,TheDAMPE:2017dtc}.
The spectral softening may be due to the breakdown of the conventional assumption of continuous source distribution or the maximum acceleration limits of
electron sources, while the peak can be explained by dark matter (DM) annihilation in a nearby clump halo either into exclusive
$e^+e^-$ final state or equally into  $e^+ e^-$, $\mu^+\mu^-$ and $\tau^+ \tau^-$ states \cite{Yuan2017}.
The best fit values for the DM particle mass, annihilation cross section, the DM halo mass and the DM annihilation luminosity $\mathcal{L}=\int \rho^2 dV$
are about 1.5 TeV, $\langle \sigma v \rangle \sim 10^{-26} \,cm^3/s$, $10^{7-8} \, M_{\rm sun}$ and $10^{64-66}\, {\rm GeV^2\,cm^{-3}}$,
respectively, if the halo is about $0.1 \sim 0.3$ kpc away from the earth \cite{Yuan2017}.

Many simplified DM models have already been proposed to interpret the DAMPE peak \cite{Fan:2017sor, Gu:2017gle, Duan:2017pkq, Zu:2017dzm, Tang:2017lfb, Chao:2017yjg, Gu:2017bdw, Athron:2017drj, Cao:2017ydw, Duan:2017qwj, Liu:2017rgs, Huang:2017egk, Chao:2017emq, Gao:2017pym, Niu:2017hqe, Gu:2017lir, Nomura:2017ohi, Zhu:2017tvk, Ghorbani:2017cey, Cao:2017sju, Li:2017tmd, Chen:2017tva,  Jin:2017qcv, Cholis:2017ccs, Fang:2017tvj, Ding:2017jdr, Yang:2017cjm, Ge:2017xyz, Liu:2017abc,Zhao:2017nrt,
Sui:2017qra,Okada:2017pgr,Profumo:2017obk}. Some are based on typical new gauged anomaly-free family $U(1)$ symmetry with the corresponding gauge boson
as the mediator \cite{Fan:2017sor,Chao:2017yjg,Cao:2017ydw,Duan:2017qwj,Chao:2017emq,Zhu:2017tvk,Cao:2017sju}. Other proposals, such as scalar mediator with
typical lepton-specific Yukawa couplings, have also been discussed \cite{Nomura:2017ohi,Li:2017tmd,Chen:2017tva,Ding:2017jdr,Liu:2017abc,Zhao:2017nrt,Sui:2017qra}.
However, many existing DM explanations of the DAMPE results are rather {\it ad hoc} and the involved interactions are not naturally
the consequence of a well-motivated popular BSM (beyond-Standard Model) model. So it is desirable to see if some popular BSM models can already explain the DAMPE results.

The minimal LRSM predicts 14 physical Higgs bosons: four CP-even  $H_{1,2,3,4}$, two CP-odd $A_{1,2}$, four singly charged $H^\pm_{1,2}$ as well as four doubly charged $H^{\pm\pm}_{1,2}$, all with increasing masses in the ascending order. In this work we propose to explain the DAMPE excess with the complex scalar DM annihilation into triplet scalar pairs which later decay and produce cosmic leptons. More specifically, we will consider the following two scenarios:
\begin{itemize}
\item Scenario-I: $\chi\chi^* \to H_1^{++}H_1^{--} \to \ell_i^+\ell_i^+\ell_j^-\ell_j^-$,
\item Scenario-II: $\chi\chi^* \to H_{k}^{++} H_{k}^{--} \to \ell_i^+\ell_i^+\ell_j^-\ell_j^-$ accompanied by $\chi\chi^* \to H_1^+ H_1^- \to \ell_i^+ \nu_{\ell_i} \ell_j^- \nu_{\ell_j}$,
\end{itemize}
where $\chi$ stands for the scalar DM candidate, $\ell_{i,j}=e,\mu,\tau$ and $k=1,2$.

As for the proposal, we stress that the mediating scalars $H_{1,2}^{\pm \pm}$ and $H_1^\pm$
can naturally arise from the LRSM where they belong to the $SU(2)_L$ triplet $\Delta_L$ and/or the $SU(2)_R$ triplet $\Delta_R$, which are essential in generating
neutrino masses. With certain assumptions on the form of the Dirac mass terms for neutrinos, the Yukawa couplings involving the triplet
scalars can be nearly generation universal for leptons or first generation dominated over the other generations. As a result, the scalars
can decay democratically into three generation of leptons or dominantly into electrons.

We also stress that the DM can be an
intrinsic component of the LRSM with the DM stability guaranteed by either the minimal dark matter spirit or due to matter parity \cite{Heeck:2015qra}.
In the former case, the DM particle can be identified as the neutral component within certain high-dimensional SU(2) representations
that forbids the renormalizable couplings leading to its decay. In the latter case, however, the residue $Z_2^{B-L}$ symmetry from the
$U(1)_{B-L}$ breaking by the scalar triplet Higgs $\Delta_{L,R}$ can also act as the DM parity, which could guarantee the stability of
alternative fermionic (bosonic) DM candidates with even (odd) $B-L$ charge \cite{Heeck:2015qra,Garcia-Cely:2015quu}.

This paper is organized as follows. In Section~\ref{Section-2}, we give a brief review about the essential features of the minimal LRSM.
In Section~\ref{Section-3}, we propose a simplified scalar DM theory, which is based on the LRSM, to explain the DAMPE excess.
In Section~\ref{Section-4}, we briefly discuss the implication of our explanation and its test at colliders. Finally, we draw our conclusions in Section~\ref{Section-5}.

\section{\label{Section-2}Brief review of the minimal LRSM}

As noted previously, the LRSM model is an extension of the SM with the  corresponding gauge group $SU(3)_C \times SU(2)_L \times SU(2)_R \times U(1)_{B-L}$, and
all the right-handed fermions are embedded into the $SU(2)_R$ doublets. Due to such an assignment, right-handed neutrinos, which are needed to fit into
right-handed lepton doublets, naturally appear in LRSM.

In the minimal LRSM, the quantum numbers of the particle contents under $SU(3)_c\times SU(2)_L\times SU(2)_R \times U(1)_{B-L}$ are given by \cite{Bonilla:2016fqd}
\begin{subequations}
\begin{align}
\text{Fermions:}\quad& \notag\\
Q_L &= \begin{pmatrix}
u_L \\ d_L
\end{pmatrix} \in (\mathbf{3},\mathbf{2},\mathbf{1},1/3)\,, & Q_R &= \begin{pmatrix}
u_R \\ d_R
\end{pmatrix} \in (\mathbf{3},\mathbf{1},\mathbf{2},1/3)\,, \\
L_L &= \begin{pmatrix}
\nu_L \\ \ell_L
\end{pmatrix} \in (\mathbf{1},\mathbf{2},\mathbf{1},-1)\,, & L_R &= \begin{pmatrix}
\nu_R \\ \ell_R
\end{pmatrix} \in (\mathbf{1},\mathbf{1},\mathbf{2},-1)\,.
\end{align}
\begin{align}
\text{Scalars:}\quad& \notag\\
\Phi &=\begin{pmatrix}
\phi^{0}_{1}   &   \phi^{+}_{1} \\
\phi^{-}_{2}   &   \phi^{0}_{2}
\end{pmatrix} \in  (\mathbf{1},\mathbf{2},\mathbf{2},0) \,, \\
\Delta_{L} &=\begin{pmatrix}
\frac{\delta^{+}_{L}}{\sqrt{2}}& \delta^{++}_{L} \\
\delta^{0}_{L}                 & -\frac{\delta^{+}_{L}}{\sqrt{2}}
\end{pmatrix}\in (\mathbf{1},\mathbf{3},\mathbf{1},2)\,,
& \Delta_{R} &=\begin{pmatrix}
\frac{\delta^{+}_{R}}{\sqrt{2}}& \delta^{++}_{R} \\
\delta^{0}_{R}                 & -\frac{\delta^{+}_{R}}{\sqrt{2}}
\end{pmatrix}\in (\mathbf{1},\mathbf{1},\mathbf{3},2)\,,\nn
\end{align}
\end{subequations}
where the Bi-doublet Higgs field is needed to give masses to ordinary SM fermions other than neutrinos, and
the triplet fields are needed to generate tiny neutrino masses via mixed Type-I and Type-II seesaw mechanisms and meanwhile
preserve the left-right symmetry.

The Lagrangian is as follows \cite{Bonilla:2016fqd}
\begin{eqnarray}
\mathcal{L} =\mathcal{L}_{kin}+ \mathcal{L}_Y^\Phi + \mathcal L_Y^{\Delta} + \mathcal L_{LR}~,
\end{eqnarray}
where the Yukawa couplings involving the bi-doublets and triplets scalars are given by
\begin{eqnarray} \label{eq:model:dirac_masses}
  -\mathcal{L}_Y^\Phi &= & \overline{Q_L}\left(Y_{Q_1}\Phi +Y_{Q_2} \tilde{\Phi}\right) Q_{R} +
   \overline{L_L}\left(Y_{L_1}\Phi +Y_{L_2} \tilde{\Phi}\right) L_{R} + {\rm h.c.}\,, \nonumber \\
 -\mathcal L_Y^{\Delta} &=& \overline{L_L^C} \,Y_{\Delta_{L}}\,(i\sigma_2) \Delta_L\, L_L
                             +\overline{L_R^C} \, Y_{\Delta_{R}}\, (i\sigma_2) \Delta_R\, L_R
                + {\rm h.c.}\,,
\end{eqnarray}
with  $\tilde{\Phi}\equiv -\sigma_{2}\Phi^{\ast}\sigma_{2}$ and $\overline{\Psi^C} = i \Psi^T  \gamma_2 \gamma_0$,
and the Higgs potential takes following form
\begin{eqnarray}
\mathcal{L}_{LR} &=& - \mu_1^2 {\rm Tr} (\Phi^{\dag} \Phi) - \mu_2^2
\left[ {\rm Tr} (\tilde{\Phi} \Phi^{\dag}) + {\rm Tr} (\tilde{\Phi}^{\dag} \Phi) \right]
- \mu_3^2 \left[ {\rm Tr} (\Delta_L \Delta_L^{\dag}) + {\rm Tr} (\Delta_R
\Delta_R^{\dag}) \right] 
\\
&&+ \lambda_1 \left[ {\rm Tr} (\Phi^{\dag} \Phi) \right]^2 + \lambda_2 \left\{ \left[
{\rm Tr} (\tilde{\Phi} \Phi^{\dag}) \right]^2 + \left[ {\rm Tr}
(\tilde{\Phi}^{\dag} \Phi) \right]^2 \right\} \nonumber \\
&&+ \lambda_3 {\rm Tr} (\tilde{\Phi} \Phi^{\dag}) {\rm Tr} (\tilde{\Phi}^{\dag} \Phi) +
\lambda_4 {\rm Tr} (\Phi^{\dag} \Phi) \left[ {\rm Tr} (\tilde{\Phi} \Phi^{\dag}) + {\rm
Tr}
(\tilde{\Phi}^{\dag} \Phi) \right]\nonumber \\
&& + \rho_1 \left\{ \left[ {\rm Tr} (\Delta_L \Delta_L^{\dag}) \right]^2 + \left[ {\rm
Tr} (\Delta_R \Delta_R^{\dag}) \right]^2 \right\} \nonumber \\ && + \rho_2 \left[ {\rm
Tr} (\Delta_L \Delta_L) {\rm Tr} (\Delta_L^{\dag} \Delta_L^{\dag}) + {\rm Tr} (\Delta_R
\Delta_R) {\rm Tr} (\Delta_R^{\dag} \Delta_R^{\dag}) \right] \nonumber
\\
&&+ \rho_3 {\rm Tr} (\Delta_L \Delta_L^{\dag}) {\rm Tr} (\Delta_R \Delta_R^{\dag})+
\rho_4 \left[ {\rm Tr} (\Delta_L \Delta_L) {\rm Tr} (\Delta_R^{\dag} \Delta_R^{\dag}) +
{\rm Tr} (\Delta_L^{\dag} \Delta_L^{\dag}) {\rm Tr} (\Delta_R
\Delta_R) \right]  \nonumber \\
&&+ \alpha_1 {\rm Tr} (\Phi^{\dag} \Phi) \left[ {\rm Tr} (\Delta_L \Delta_L^{\dag}) +
{\rm Tr} (\Delta_R \Delta_R^{\dag})  \right] \nonumber
\\
&&+ \left\{ \alpha_2 e^{i \delta_2} \left[ {\rm Tr} (\tilde{\Phi} \Phi^{\dag}) {\rm Tr}
(\Delta_L \Delta_L^{\dag}) + {\rm Tr} (\tilde{\Phi}^{\dag} \Phi) {\rm Tr} (\Delta_R
\Delta_R^{\dag}) \right] + {\rm h.c.}\right\} \nonumber
\\
&&+ \alpha_3 \left[ {\rm Tr}(\Phi \Phi^{\dag} \Delta_L \Delta_L^{\dag}) + {\rm
Tr}(\Phi^{\dag} \Phi \Delta_R \Delta_R^{\dag}) \right] + \beta_1 \left[ {\rm Tr}(\Phi
\Delta_R \Phi^{\dag} \Delta_L^{\dag}) +
{\rm Tr}(\Phi^{\dag} \Delta_L \Phi \Delta_R^{\dag}) \right] \nonumber \\
&&+ \beta_2 \left[ {\rm Tr}(\tilde{\Phi} \Delta_R \Phi^{\dag} \Delta_L^{\dag}) + {\rm
Tr}(\tilde{\Phi}^{\dag} \Delta_L \Phi \Delta_R^{\dag}) \right] + \beta_3 \left[ {\rm
Tr}(\Phi \Delta_R \tilde{\Phi}^{\dag} \Delta_L^{\dag}) + {\rm Tr}(\Phi^{\dag} \Delta_L
\tilde{\Phi} \Delta_R^{\dag}) \right] \,. \nonumber
\end{eqnarray}
In above potential, $\mu_i$, $\lambda_i$, $\beta_i$ with $i=1,2,3$ and $\rho_j$, $\alpha_j$ with $j=1,\cdots,4$ are all free parameters.

The $SU(2)_R\tm U(1)_{B-L}$ is broken to $U(1)_Y$ by the VEV of the $SU(2)_R$ triplet scalar $\Delta_R$, while the SM gauge group
is broken to $U(1)_Q$ by the VEVs of the bi-doublet Higgs. The VEVs of the bi-doublet and triplets, which are taken to be
real to forbid spontaneously CP violation, are parameterized as
\begin{eqnarray}
\langle\phi_1^0\rangle &=& \frac{v}{\sqrt{2}} \cos \beta\,,\qquad \langle\phi_2^0\rangle = \frac{v}{\sqrt{2}} \sin\beta\,, \qquad t_\beta \equiv \tan\beta=\frac{v_2}{v_1}\,,\nn\\
\langle\delta_L^0\rangle &=& \frac{v_L}{\sqrt{2}}, ~~~~~~ \qquad \langle\delta_R^0\rangle = \frac{v_R}{\sqrt{2}}\,,
\end{eqnarray}
with $v_L\ll v \ll v_R$, so $v$ can be identified as the SM VEV. The masses of the new gauge bosons therefore read
\begin{eqnarray}
M_{Z_R} \simeq \sqrt{g_R^2 + g_{BL}^2}\, v_R\,, \qquad\qquad M_{W_R} \simeq \frac{g_R}{\sqrt{2}} \, v_R \,.
\end{eqnarray}
Due to the LR symmetry, we take the two $SU(2)$ gauge coupling to be equal, namely $g_R=g_L$.
 The mixing between the electric charged gauge fields $W_{L}$ and $W_{R}$ will result in two mass eigenstates $W$ and $W^\prime$, and similarly, the mixing
 among the neutral components $W_L^3,W_R^3,B_{B-L}$ will predict three vector bosons as mass eigenstates, i.e. photon, $Z$ and $Z^\prime$.

The minimal LRSM predicts ten physical particles: four CP-even Higgs bosons, two CP-odd Higgs bosons, two singly charged Higgs bosons as well as two doubly charged
Higgs bosons. With the minimization conditions of the scalar potential, one can trade the parameters $\mu_i$ and $\beta_2$ by the vacuum expectation values (VEV) \cite{Bonilla:2016fqd}. As a result, with the assumption $v_L/v, v_L/v_R, \tan \beta, \alpha_1, \alpha_2, \beta_1 \to 0$ (so that the mixings between the Bi-doublet and the triplet
scalars are small),  the Bidoublet-like scalar masses are given by
\begin{subequations}
\begin{align}
m_h^2 &\simeq 2 \lambda_1 v^2-\frac{8 \lambda_4^2 v^4}{\alpha_3 v_R^2}\,, &
m_H^2 &\simeq 2 (2\lambda_2+\lambda_3)v^2 + \frac{\alpha_3}{2} v_R^2\,,  \label{eq:bidoubletHiggs_masses}  \\
m_{A}^2 &\simeq  \frac{\alpha_3}{2} v_R^2 + 2(\lambda_3 - 2 \lambda_2)  v^2 \,, &
m_{H^{\pm}}^2&\simeq \frac{1}{4}\alpha_3(v^2+2v_R^2)\,,
\end{align}
\end{subequations}
where $h$ corresponds to the SM-like Higgs boson with its mass fixed at $125 {\rm GeV}$; $H, A$ and $H^\pm$ are the heavier neutral scalar and pseudoscalar states as well as the charged Higgs, respectively. The triplet-scalar sector masses are:
\begin{subequations}
\begin{align}
m_{H_L}^2 &\simeq \frac{1}{2} \left(\rho_3 - 2 \rho_1\right) v_R^2\, &
m_{H_R}^2 &\simeq 2 \rho_1 v_R^2\,,  \\
m_{A_{L}}^2 &\simeq \frac{1}{2}(\rho_3 - 2\rho_1) v_R^2\,, &
m_{H^{\pm}_L}^2&\simeq \frac{1}{2} (\rho_3 -2 \rho_1) v_R^2 \,, \\
m_{H^{\pm\pm}_a}^2&\simeq 2 \rho_2 v_R^2 +\frac{1}{2} \alpha_3 v^2\,, &
m_{H^{\pm\pm}_b}^2&\simeq \frac{1}{2} \left((\rho_3-2\rho_1)v_R^2 + \alpha_3 v^2 \right)\,,
\end{align}
\end{subequations}
where particles with an index $L(R)$ mostly consist of $\Delta_{L(R)}$ components, and since the doubly-charged Higgses can in general be strongly mixed,
we label them as $H^{\pm\pm}_{a/b}$. Under the condition $v_L/v_R \to 0$, the mass spectrums implies following approximate degeneracies
\begin{eqnarray}
\label{eq-bi-1}2 \lambda_1 v^2-\frac{8 \lambda_4^2 v^4}{\alpha_3 v_R^2} &\simeq& m_h^2,\\
\label{eq-bi-2}\frac{\alpha_3}{2} v_R^2 &\simeq& m_H^2 \simeq m_A^2 \simeq m_{H^\pm}^2
\end{eqnarray}
for the bi-doublet sector and
\begin{eqnarray}
\label{eq-tri-1}2 \rho_1 v_R^2 &\simeq& m_{H_R}^2, \\
\label{eq-tri-2}2 \rho_2 v_R^2 +\frac{1}{2} \alpha_3 v^2 &\simeq& m_{H^{\pm\pm}_a}^2, \\
\label{eq-tri-3}\frac{\rho_3-2\rho_1}{2} v_R^2 &\simeq& m_{H^{\pm\pm}_b}^2 \simeq m_{H_L^\pm}^2 \simeq m_{H_L}^2 \simeq m_{A_L}^2
\end{eqnarray}
for the triplet sector. In the following, we label particles with the same quantum numbers by the subscripts $1, 2, \cdots$ and assume that they have an ascending mass order, i.e.
\begin{eqnarray}
\label{eq-H-1234} m_{H_{1,2,3,4}} &:& \quad \text{neutral CP-even Higgs},\\
\label{eq-A-12} m_{A_{1,2}} &:& \quad \text{neutral CP-odd Higgs},\\
\label{eq-Hpm-12} m_{H^\pm_{1,2}} &:& \quad \text{singly charged Higgs},\\
\label{eq-Hpmpm-12} m_{H^{\pm\pm}_{1,2}} &:& \quad \text{doubly charged Higgs.}
\end{eqnarray}

In the minimal LRSM, the tiny neutrino mass can be generated via mixed Type-I and Type-II seesaw mechanisms with the corresponding mass matrix given by
\cite{Bonilla:2016fqd}
\begin{align}
\label{eq:neutrino_mass_matrix}
\frac{1}{2} \begin{pmatrix}\overline{\nu_L} & \overline{\nu_R^C}\end{pmatrix} \,
\begin{pmatrix}
   M_L^* &   M_D\\
   M_D^T         &   M_R
\end{pmatrix}\begin{pmatrix}\nu_L^C\\ \nu_R \end{pmatrix} \,+\, {\rm h.c.}\,,
\end{align}
where
\begin{align}\label{eq:ML_MR_mD_defs}
 M_{L}=\sqrt{2}\,Y_{\Delta_{L}}v_{L}\,,\ \
 M_{R}=\sqrt{2}\,Y_{\Delta_{R}}v_{R}\,,\ \ \text{and}\ \
 M_D=\frac{v}{\sqrt{2}}\left(Y_{L_{1}}\sin\beta + Y_{L_{2}}\cos\beta\right)\,.
\end{align}
After the diagonalization of the mass matrix, the Majorana mass of the left-handed neutrinos can be determined to be
\begin{eqnarray} \label{eq:lightnumassmatrix}
m_{\nu}^{\rm light}&=\left(M_L^*-M_D\,M_{R}^{-1}\,M_D^T\right)\,.
\end{eqnarray}
This expression indicates that possibly large cancelation among the two terms is required to give tiny neutrino masses of order 0.1 eV.  One
should note that loop corrections will in general spoil the cancelation among the two terms.

The discrete LR symmetry, which can be identified with Parity symmetry, requires the Yukawa couplings to satisfy  \cite{Bonilla:2016fqd}
\beqa
Y_{a}=Y_a^\da~,~~~~~Y_{\Delta_L}=Y_{\Delta_R}~.
\eeqa
In terms of the neutrino masses and PMNS mixing matrix from neutrino oscillation experiments, the Yukawa couplings involving the
triplets can be determined as  \cite{Bonilla:2016fqd}
\begin{eqnarray}
&&Y_\Delta^{(\pm\pm\pm)} \equiv  Y_{\Delta_{L/R}}^{(\pm\pm\pm)} = \frac{1}{{2\sqrt{2} v_L}} M_D^{* 1/2} R^* \text{diag}\left(\BDi \pm \sqrt{\left(\BDi\right)^2 + 4 \alpha}\right)
R^\dag M_D^{1/2}\,,\nn\\
&&B_D = R^{\dagger} M_D^{* -1/2} m_{\nu}^{\rm light}  M_D^{\,-1/2} R^*~,~~~~\alpha=v_L/v_R.
\label{eq:TripletCouplSolution}
\end{eqnarray}
for a given specific input of $v_L$, $v_R$ and $M_D$. Within the previous expression, $B_D$ is a diagonal $3\times3$ matrix and $R$ is a unitary rotation matrix
to keep both sides of (\ref{eq:TripletCouplSolution}) equal. It should be noted that there does not exist a unique solution to the triplet-Yukawa couplings, which corresponds
to an ambiguous $\pm$ sign in the bracket of the expression. It should also be noted that the magnitude of the $Y_\Delta$ is sensitive to the choice of $v_L$ and $M_D$, and may vary from ${\cal{O}}(10^{-4})$ to ${\cal{O}} (1)$ in producing the measured neutrino masses and mixings.

\section{\label{Section-3}DAMPE explanation with scalar DM}

\begin{table}
\begin{center}
\begin{tabular}{lr@{\hspace{5em}}lr}\toprule
$v_L$ 		& $2.0 \times 10^{-7}$\, {\rm GeV}		& $v_R$				& $2.0 \times 10^4$\, {\rm GeV}	\\
$M_D$  		& $\mathbb{1}\, {\rm MeV}$	 	& $Y_\Delta^{(+++)}$	& Eq.(\ref{eq:Ydelta_diag_ppp})  \\
$m_\chi$ 		& (2.95, 3.15) {\rm TeV}	 		&  $\tan\beta$	& $10^{-4}$  \\
$\lambda_1$	&  0.13				& $\lambda_2$	& 0 \\
$\lambda_3$ 	&  0  					& $\lambda_4$	& 0 \\
$\rho_1$		&  (0,\,0.1)				& $\rho_2$   	& (0,\,0.1)	 \\
$\rho_3$ 		&  (0,\,0.2)				& $\rho_4$   	& 0 \\
$\alpha_1$	&  0					& $\alpha_2$	& 0 \\
$\alpha_3$ 	&  2.0 			& $\beta_1$	& 0 \\
$\beta_3$ 	&  0		 			& $\kappa_1$	& (0,\,5) \\
$\kappa_{2,3}$ 	&  0 					& $\lambda_\chi$	& 0 \\
\bottomrule
\end{tabular}
\caption{Parameter settings for the scan in this work over the varying parameters. These parameters are defined at the scale of $v_R = 20 \,{\rm TeV}$.}
\label{Table-scan}
\end{center}
\end{table}

\begin{figure}[tbp]
\begin{center}
\includegraphics[height=6cm,width=8cm]{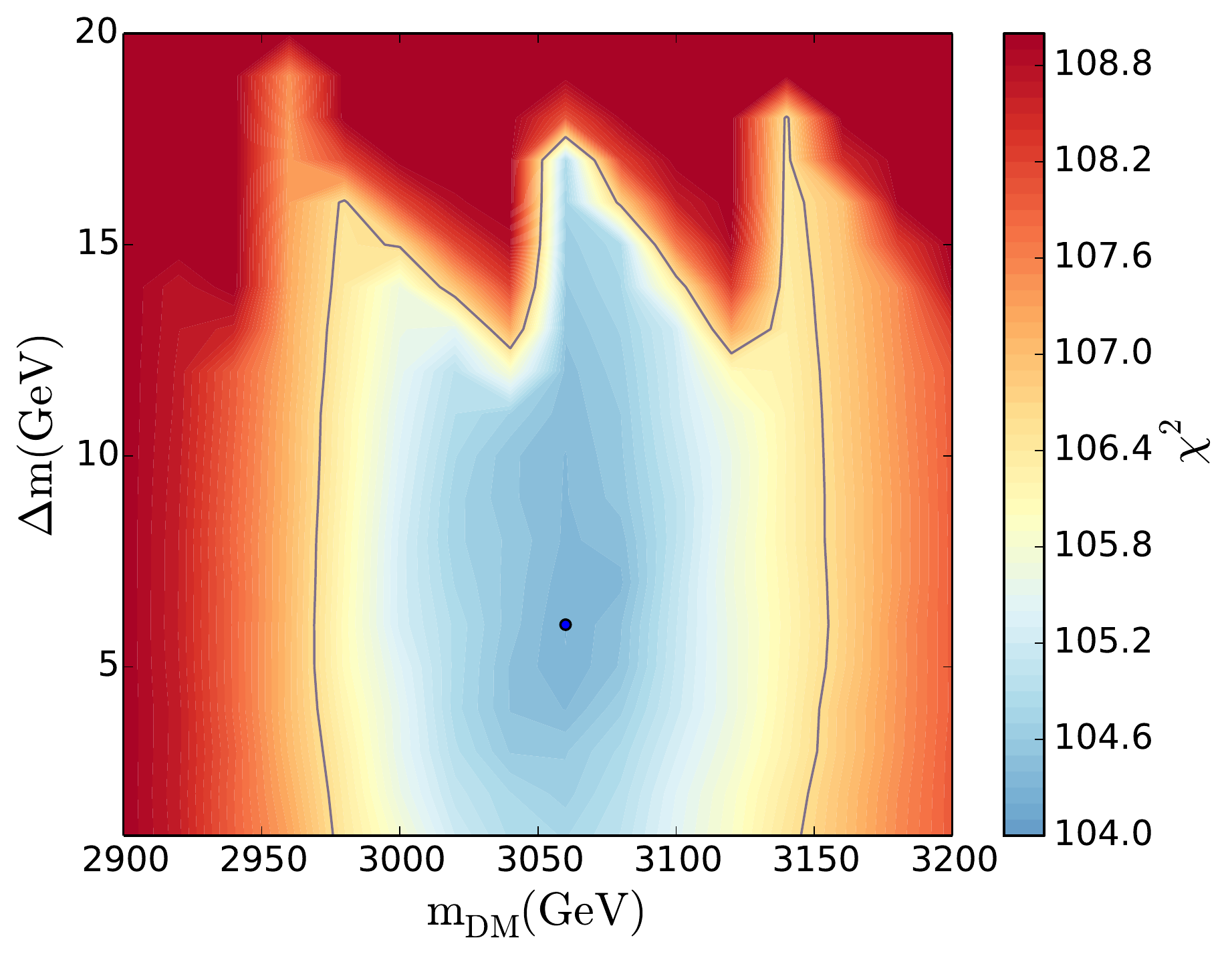}
\includegraphics[height=6cm,width=6cm]{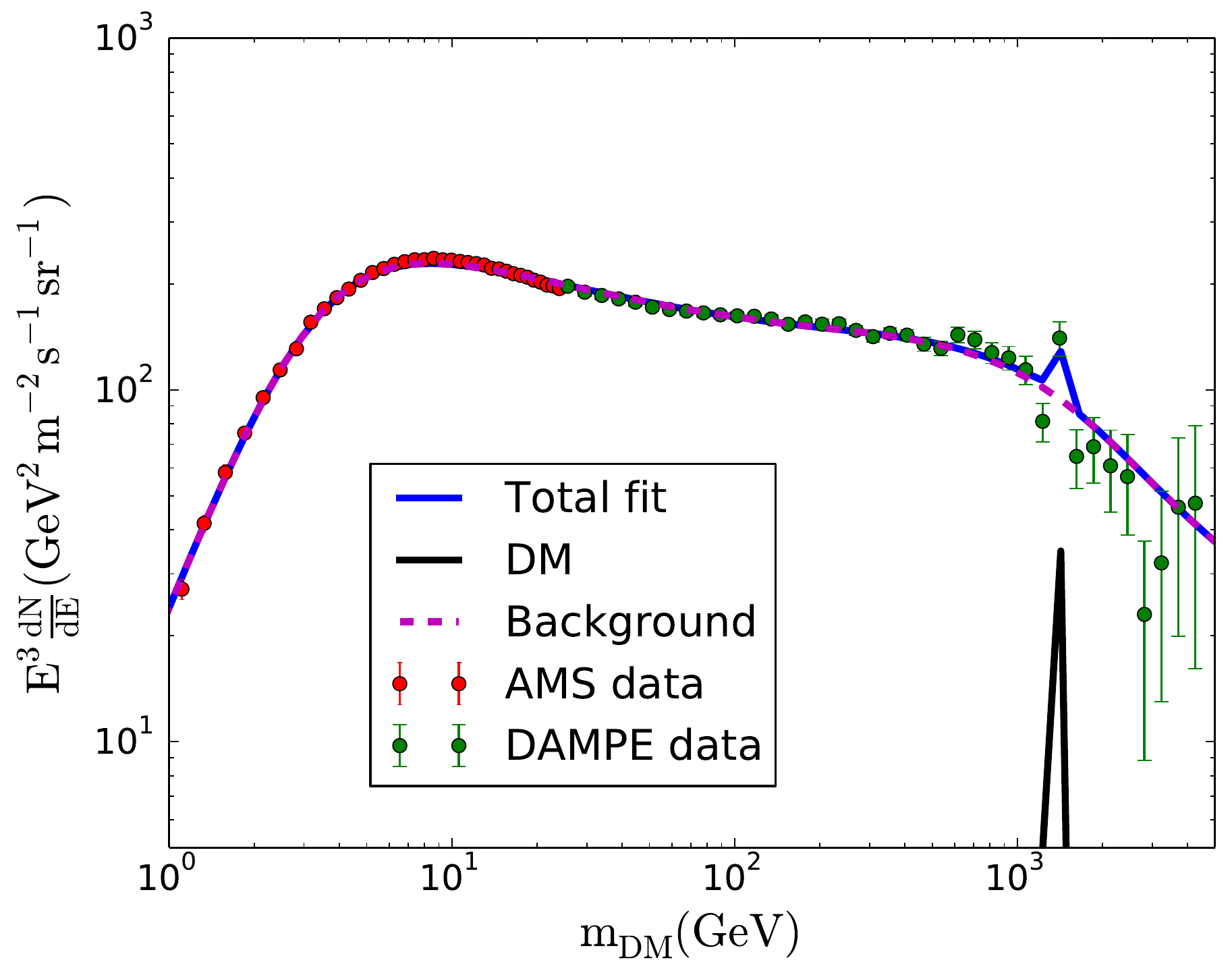}
\caption{Fit of the $e^+e^-$ spectrum generated by the process  $\chi \chi \to H^{++}_1 H^{--}_1$ with  $H^{\pm \pm}_1 \to e^\pm e^\pm, \mu^\pm \mu^\pm, \tau^\pm \tau^\pm $
at equal rate to the AMS-02 and DAMPE data. \text{\bf Left} panel: $\chi^2$ map projected on $\Delta m-m_\chi$ plane with $\Delta m \equiv  m_\chi - m_{H^{\pm \pm}_1}$
and the color bar denoting the $\chi^2$ values. The best fit point locates at about $(6 \, {\rm GeV}, 3060 \, {\rm GeV})$, and the contour of $\chi^2 = \chi^2_{best} + 2.3$ (solid line) is also plotted.
\text{\bf Right} panel:  The cosmic $e^+e^-$ spectrum of the best fit point generated by the DM annihilation process in comparison with the AMS-02 and DAMPE data. }
\label{Fit-spectrum}
\end{center}
\end{figure}

As one of the most compelling BSM theories, the LRSM itself can naturally accommodate a DM candidate, which is absolutely stable
due to the residual matter parity $Z_2^{B-L}$ \cite{Heeck:2015qra,Garcia-Cely:2015quu}. With respect to a scalar DM candidate, the minimal
realization requires the introduction of two complex scalar fields $\phi_L$ and $\phi_R$ with their respective quantum numbers
under the gauge group $SU(3)_c\times SU(2)_L\times SU(2)_R \times U(1)_{B-L}$ \cite{Garcia-Cely:2015quu}
\begin{subequations}
\begin{align}
\phi_{L} &=\begin{pmatrix}
\phi_L^0 \\
\phi_L^-
\end{pmatrix}\in (\mathbf{1},\mathbf{2},\mathbf{1},\mathbf{-1})\,,
& \phi_{R} &=\begin{pmatrix}
\phi_R^0 \\
\phi_R^-
\end{pmatrix}\in (\mathbf{1},\mathbf{1},\mathbf{2},\mathbf{-1})\,, \nn
\end{align}
\end{subequations}
and the DM candidate corresponds to the lightest mass eigenstate among the mixing of the neutral components\footnote{Note that for such an assignment of gauge quantum numbers, the properties of the fields $\phi_L$
and $\phi_R$ are similar to those of the left-handed and right-handed scalar lepton fields in supersymmetric LRSM, respectively. Thus in some cases the DM as the lightest state
can mimic the behavior of the popular sneutrino DM in supersymmetric theories.}.  These additional gauge non-singlet scalars extend greatly
the minimal LRSM, and the general form of the resulting theory contains 16 new parameters, and twenty extra quartic scalar interactions \cite{Garcia-Cely:2015quu}.
Obviously, working in such a complex framework to interpret the DAMPE excess involves the treatment of a large number of free parameters,
which usually obscures the underlying physics. This situation motivates us to consider the incorporation of DM physics into the LRSM in a simpler way.
To be more specific, we note that
the gauge singlet scalar is ubiquitous in various UV completion theories of the LRSM. For example, it can appear in the decomposition of {\bf 54} or {\bf 210}
dimensional representation of SO(10) under Pati-Salam $SU(4)_c\tm SU(2)_L\tm SU(2)_R$ (or LR gauge group) gauge group and possibly be tuned to be light \cite{Slansky:1981yr}.
It can also appear in orbifold GUT models with proper boundary conditions by embedding the LRSM gauge group into GUT or partial unification theory
\cite{Kawamura:1999nj,Hall:2001pg}. These facts motivate us to consider a VEV-less gauge singlet complex scalar particle as the DM candidate with its
stability protected by an accidental global $U(1)_\chi$ symmetry, which may be promoted to a gauged one that was broken at a certain high energy scale into a discrete $Z_2$ symmetry
in the early evolution of the Universe.

Based on the above discussion,  we propose to introduce a complex scalar DM $\chi$ into the minimal LRSM,
which is a singlet under the LR gauge group,
as the simplest effective DM model to explain the DAMPE results with the DM stability guaranteed by the conserved accidental $U(1)_\chi$ quantum number, 
just as the conserved baryon number ensures the stability of proton classically.
The couplings of $\chi$ are assumed to be
 \beqa
{\cal L}&\supseteq& |\pa_\mu \chi|^2 -\mu_\chi^2 |\chi|^2 -\kappa_1 |\chi|^2  \left[ {\rm Tr} (\Delta_L \Delta_L^{\dag})
 + {\rm Tr} (\Delta_R
\Delta_R^{\dag}) \right]-\kappa_2 |\chi|^2 {\rm Tr} (\Phi^{\dag} \Phi)~\nn\\
&&-\kappa_3 |\chi|^2 \left[ {\rm Tr} (\tilde{\Phi} \Phi^{\dag}) + {\rm Tr} (\tilde{\Phi}^{\dag} \Phi) \right] - \lambda_\chi | \chi |^4
~,
\label{eq-L-ours}
\eeqa
which are invariant under the discrete Left-Right symmetry for real parameters $\lambda_\chi$ and $\kappa_i$ (i=1,2,3).

We use the package \textbf{SARAH} \cite{Staub:2015kfa} to implement the model and the package \textbf{SPheno} \cite{Porod:2003um,Porod:2011nf} to calculate the mass spectrum.
Since a lot of parameters are involved in our discussion on the interpretation of the excess, we fix some of them in Table \ref{Table-scan},
where $v_L=2 \times 10^{-7} \,{\rm GeV}$, $M_D = \mathbb{1}\,{\rm MeV}$ and $"+++"$ sign choices for three generations will give
\begin{equation}
Y_\Delta^{(+++)} = \begin{pmatrix}
\SI{1.12E-2}{}  & \SI{-1.41E-5}{}   & \SI{2.97E-6}{}  \\
\SI{-1.41E-5}{} & \SI{1.12E-2}{}   & -\SI{3.78E-5}{}   \\
\SI{2.97E-6}{} & -\SI{3.78E-5}{} & \SI{1.12E-2}{}
\end{pmatrix}\,.
\label{eq:Ydelta_diag_ppp}
\end{equation}
according to Eq.(\ref{eq:TripletCouplSolution}). This setup, where the Dirac neutrino mass is diagonal and flavor-universal, always predicts an almost degenerate spectrum of right-handed neutrinos due to the nearly degenerate diagonal $Y_\Delta^{(i,i)}$ entries and meanwhile the relatively smaller non-diagonal entries. As a result, the triplet-dominated scalars $H_{1,2}^{\pm \pm}$ will decay dominantly into $l^\pm_i l^\pm_i$ ($i= e, \mu, \tau$) with approximately equal branching ratios.

\begin{table}
\begin{center}
\begin{tabular}{|c|c|c|}
\hline
Scenario & \makecell{Mass spectrum:  $m_\chi$ and $\Delta m > 0$ lie within \\
the region enclosed by the solid line in Fig.\ref{Fit-spectrum}. } & \makecell{Relevant DM annihilations \\ $\ell_{i,j}=e,\mu,\tau$,\, $k=1,2$. } \\
\hline\hline
\textbf{I} 			& \makecell{$m_{H_1^{\pm\pm}} \sim 3 \,{\rm TeV}$ and $\Delta m \equiv m_\chi - m_{H_1^{\pm \pm}} $. \\ The other scalars are heavier than DM.}	
		 & \makecell{ $\chi\chi \to H_1^{++}H_1^{--} \to \ell_i^+\ell_i^+\ell_j^-\ell_j^-$ }\\
\hline
\textbf{II} 			& \makecell{$m_\chi > m_{H_{2}^{\pm\pm}},\, m_{H_2},\, m_{A_1},\, m_{H^\pm_1} >  m_{H_{1}^{\pm\pm}} $, \\
and $\Delta m \equiv m_\chi - m_{H_1^{\pm \pm}}$.  \\ The other scalars are heavier than DM.}	& \makecell{$\chi\chi \to H_{k}^{++}H_{k}^{--} \to \ell_i^+\ell_i^+\ell_j^-\ell_j^-$ \\ $\chi\chi \to H_1^+ H_1^- \to \ell_i^+ \nu_{\ell_i} \ell_j^- \nu_{\ell_j}$ }\\
\hline
\end{tabular}
\caption{Two scenarios of Higgs spectrum and DM annihilation channels pertinent to explain the DAMPE excess.}
\label{Table-scenario}
\end{center}
\end{table}

In practice, in order to obtain the solutions to the DAMPE excess we first determine the favored DM mass $m_{\chi}$
and $\Delta m \equiv m_\chi - m_{H_1^{\pm \pm}}$ when we utilize the process $\chi\chi \to H_1^{++}H_1^{--} \to \ell_i^+\ell_i^+\ell_j^-\ell_j^-$
with $\ell_{i,j}=e,\mu,\tau$ to generate the measured $e^+ e^-$ spectrum. The impact of the mass spectrum on the $e^+e^-$ flux and also
our strategy to get their favored region have been described in detail in \cite{Cao:2017sju}. Here we simply apply them to the case in which the intermediate scalars
$H^{\pm \pm}_1$ as the DM direct annihilation products decay democratically into $e^\pm e^\pm$, $\mu^\pm \mu^\pm$ and $\tau^\pm \tau^\pm$.
The results are presented in Fig.\ref{Fit-spectrum}, where we perform the fit
of the predicted $e^+e^-$ spectrum to the corresponding AMS-02 and DAMPE data. The \textbf{Left} panel is the $\chi^2$ map on the $\Delta m - m_{\chi}$ plane
with the color bar denoting the values of the $\chi^2$ and the enclosed line corresponding to the constant contour of $\chi^2 = \chi^2_{best} + 2.3$.
The region bounded by this contour is interpreted as the best region of the two-step DM annihilation process to explain the DAMPE excess
at $1 \sigma$ level. The best fit point locates at about $(6 \, {\rm GeV}, 3060 \, {\rm GeV})$ with
$\langle \sigma v \rangle_0=2.98\times 10^{-26} \,{\rm cm^3/s}$ for the default setting on the distance of the subhalo away from the earth
and the subhalo mass in \cite{Cao:2017sju}, $d = 0.1 \,{\rm kpc}$ and $M_{\rm halo}= 1.9 \times 10^{7} m_\odot$. The \textbf{Right} panel of
Fig.\ref{Fit-spectrum} corresponds to the $e^+e^-$ spectrum predicted by the best fit point which lowers the $\chi^2$ value to 104.2 in comparison with
109.7 for the background-only hypothesis. These facts indicate that, by choosing appropriate $(\Delta m, m_\chi)$, the process $\chi \chi \to H_1^{++}
H_1^{--} \to l^+ l^+ l^{\prime -} l^{\prime -}$ is indeed capable of re-producing the DAMPE $e^+e^-$ peak.

\begin{figure}[t]
\begin{center}
\includegraphics[width=10cm]{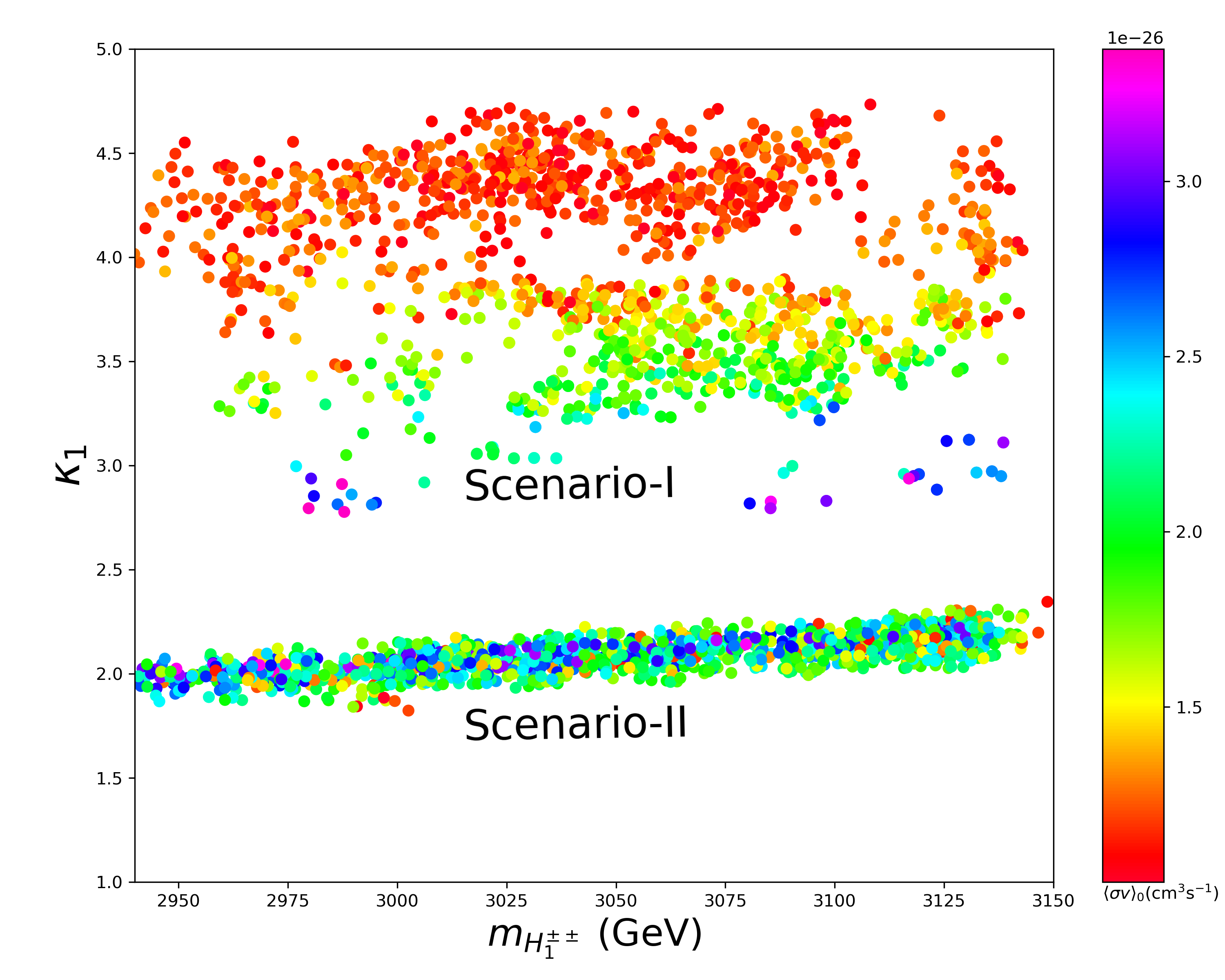}
\caption{Scenario-I and Scenario-II samples in Table \ref{Table-scenario} to explain the DAMPE data, which are projected on $\kappa_1 - m_{H_1^{\pm \pm}}$ plane with the color bar
denoting the value of $\langle \sigma v \rangle_0$. The upper and lower regions correspond to Scenario I and II, respectively, and all the samples satisfy the constraints listed in the main text. }
\label{fig-scenario-1and2-mHpmpm-kappa1}
\end{center}
\end{figure}

Next we discuss in detail two scenarios presented in Table \ref{Table-scenario} to explain the excess. In order to get relevant parameter points, we scan the
parameters $m_\chi$, $\kappa_1$ and $\rho_i$ ($i=1,2,3$) with the setting in Table \ref{Table-scan}, and we consider the following constraints:
\begin{itemize}
\item DM relic density $\Omega_{\rm DM} = 0.1199\pm 0.0027$ \cite{Planck,WMAP}, which implies that
$\langle \sigma v \rangle_{\rm FO} \sim {\cal{O}}(10^{-26}) \, {\rm cm^3/s}$ with the velocity $v \sim 0.1 \,c$ in early freeze out.

In our theoretical framework, the DM annihilation into scalar pairs $S S^\ast$ proceeds through the quartic scalar interaction $\chi \chi^\ast S S^\ast$,
$s$-change exchange of any CP-even Higgs boson and also $t$-channel exchange of $\chi$. Therefore the relic density mainly limits the coupling  strength
$\kappa_1$ for the parameter setting in Table \ref{Table-scan} and the favored spectrum in Fig.\ref{Fit-spectrum}. We use the package
\textbf{micrOMEGAs} \cite{Belanger:2014vza,Belanger:2014hqa} to obtain the density in which the threshold effects are
important when DM mass is close to intermediate particle masses \cite{PhysRevD.43.3191}. We also use the \textbf{micrOMEGAs} to calculate
the DM annihilation rate in today's Universe and the DM-nucleon scattering rate discussed below.

\item The mass spectrum presented in Fig.\ref{Fit-spectrum} as well as today's DM annihilation cross section
$\langle \sigma v \rangle_0 > 1 \times 10^{-26}\, {\rm cm^3/s}$ with $v\sim 10^{-3}\,c$ in the nearby subhalo, which are essential conditions
 to explain the DAMPE result (see \cite{Yuan2017,Cao:2017sju} for more details). Obviously, the former condition limits the ranges of
 $m_\chi$ and $\rho_i$, while the latter condition as an useful supplement to DM relic density has non-trivial requirements on $\kappa_1$.

\item DM direct detection bounds on spin-independent (SI) DM-nucleon scattering cross section $\sigma^{SI}_{\chi-n}$ from the recent XENON-1T \cite{Aprile:2017iyp}
  and PandaX-II experiments \cite{Cui:2017nnn}.

  In our theory, the scattering proceeds through $t$-channel exchange of any CP-even Higgs boson \cite{Berlin:2014tja}. Considering that the CP-even Higgs fields in
  the Bi-doublet $\Phi$ have no coupling with $\chi$ since we set $\kappa_2=\kappa_3=0$ in Table \ref{Table-scan}, and that the CP-even field in $\Delta_R$ which couples to
  $\chi \chi^\ast$ with the strength proportional to $\kappa_1 v_R$ has no couplings with quarks due to the
  $U(1)_{B-L}$ charge assignment, one can conclude that the scattering rate vanishes if there is no mixing between these two types of fields
  in forming mass eigenstates, i.e. the magnitude of the rate is decided by the size of the mixing.  As far as the parameter setting
  in Table \ref{Table-scan} is concerned, we checked that the mixing is less than $10^{-8}$, which results in the scattering rate less than $10^{-14} \,{\rm pb}$. Therefore although the direct detection bounds play an important role in DM physics and must be considered in explaining the excess, they actually have no constraint on our case.

\end{itemize}

\begin{table}[]
\centering
\begin{tabular}{|c|c|c|c|c|c|c|c|c|c|}
\hline
\multicolumn{2}{|c|}{}     & \multicolumn{4}{c|}{Scenario I } & \multicolumn{4}{c|}{Scenario II } \\ \hline
\multicolumn{2}{|c|}{$m_\chi $} & $\kappa_1$   & $100\rho_1 $   & $100\rho_2 $  & $100\rho_3 $  & $\kappa_1 $  & $100\rho_1 $  &$ 100\rho_2$  &$ 100\rho_3 $ \\ \hline
\multicolumn{2}{|c|}{(2.9,3.2)}   & (2.8,4.7)  & (1,5)     & (1.1,1.2)      & (8,20)      & (1.8,2.4)    & (1,3)      & (1.1,1.2)      & (6,10)     \\ \hline
\end{tabular}%
\caption{Survived parameter ranges to explain the DAMPE excess. Note that the values of $\rho_i$ ($i=1,2,3)$ in the table are scaled
by a factor of $10^2$ and the DM mass is in unit of ${\rm TeV}$. }\label{Table-parameter-surviving}
\end{table}

The surviving samples from the scan are projected on the plane
of $m_{H_1^{\pm\pm}}$ versus $\kappa_1$ in Fig.\ref{fig-scenario-1and2-mHpmpm-kappa1} with the color bar indicating the DM annihilation rate today
$\langle \sigma v \rangle_0$. This figure indicates that there exist certain parameter regions to explain the excess without conflicting the other experimental
results, which are characterized by $\kappa_1 \in (2.8,5)$ for Scenario I, $\kappa_1 \in (1.8,2.5)$ for Scenario II and $\langle \sigma v \rangle_0 \in (1,3.2) \times 10^{-26} {\rm cm^3 s^{-1}}$ for both scenarios.
In Table \ref{Table-parameter-surviving} we present more details about the relevant parameter regions. Note that $\kappa_1$ in Scenario I is significantly larger
than that in Scenario II. The underlying reason is that $\kappa_1$ is the coupling controlling the interaction strength between DM $\chi$ and Higgs triplets as indicated in Eq.(\ref{eq-L-ours}). Because of more intermediate Higgs available in the two-step DM annihilations in Scenario II (see Table \ref{Table-scenario}), a relatively low $\kappa_1$ is enough to predict the right relic density. Also note that $\kappa_1$ in Scenario I expands
a much wider range than that in Scenario II. This is because in some rare cases of Scenario I, the DM may annihilates through the resonant $H_3$ into right handed neutrinos. We checked these cases and found that, although $H_1^{++} H_1^{--}$ is still the dominant annihilation product, the contribution of the neutrino channel to the total annihilation rate may reach about $35\%$ at freeze-out temperature.

About the DM explanation of the DAMPE excess, we emphasize that it is consistent with the other direct and indirect DM experiments (see \cite{Fan:2017sor,Liu:2017rgs}
for a detailed discussion), such as the H.E.S.S. data on the annihilation $\chi \chi \to S^* S \to 4 e$ \cite{Abdallah:2016ygi,Profumo:2017obk}, the Fermi-LAT data in the direction of the dwarf spheroidal galaxies \cite{Ackermann:2015zua}, the Planck CMB data which is sensitive to energy injection to the CMB from DM annihilations \cite{Slatyer:2015jla,Slatyer:2015kla}, and the IceCube data on DM annihilation into neutrinos \cite{Aartsen:2017ulx}. It also survives the upper bounds from XENON-10 and XENON-100 experiments on the DM scattering off electron \cite{Essig:2017kqs}. Moreover, we checked that the samples in Fig.\ref{fig-scenario-1and2-mHpmpm-kappa1} also satisfy the constraints from collider experiments and some theoretical principles recently discussed in \cite{Maiezza:2016bzp,Chakrabortty:2016wkl,Maiezza:2016ybz}, which mainly limit the parameters $\rho_i$. These constraints include\footnote{Note that these limitations are obviously weak for the parameter setting in Table \ref{Table-scan}. This is because the $Y_\Delta$ we adopt is small in magnitude and nearly flavor diagonal, all scalar other than the SM-like Higgs boson as well as the new gauge bosons are heavier than about $3 {\rm TeV}$ so that their effects are decoupled, $v_L = 2 \times 10^{-7} {\rm GeV}$ is tiny, and meanwhile the quartic scalar couplings are only moderately large.}
\begin{itemize}
\item The existence of a SM-like Higgs boson ($H_1$ corresponding to our case) with mass around 125 GeV. We examined its properties with the package
 \textbf{HiggsSignals} \cite{Bechtle:2013xfa}.

\item Collider searches for extra scalars. We calculate the couplings of the non-SM-like Higgs bosons by the \textbf{SPheno} and link them to the
  package \textbf{HiggsBounds} \cite{Bechtle:2008jh,Bechtle:2011sb,Bechtle:2013wla}. We require them to be allowed by the direct search results at colliders.

\item Low energy lepton flavor violation processes considered in \cite{Bonilla:2016fqd}, which include two body decays such as $\mu \to e \gamma$, $\tau \to e \gamma$, $\tau \to \mu \gamma$ and three body decays such as $\mu \to e e e$, $\tau \to e e e$. These processes proceed at loop level and may be enhanced greatly (in comparison with their SM predictions) by large flavor non-diagonal elements of $Y_\Delta$ if the new scalars and new vector bosons running in the loops are not too heavy. In our analysis, we use the package \textbf{FlavorKit} \cite{Porod:2014xia} to calculate the rates of the processes.

\item $B_{s,d}^0-\bar{B}_{s,d}^0$ mixing. As was shown in \cite{Bertolini:2014sua}, this constraint is rather strong and it requires the masses of the heavier doublet Higgs
  to be larger than about $20 \,{\rm TeV}$. In our discussion, we satisfy the requirement by setting $\alpha_3 =2$ and $v_R = 20 \,{\rm TeV}$.

\item The precision electroweak observable $\delta \rho$, which is defined by $\delta \rho \equiv \frac{\Pi_{ZZ} (0)}{m_Z^2} - \frac{\Pi_{WW} (0)}{m_W^2} $ with
  $\Pi_{ZZ}(0)$ and $\Pi_{WW}(0)$ denoting the self energy of $Z$ boson and $W$ boson at zero momentum, respectively \cite{Peskin:1991sw,Lavoura:1993nq} \footnote{Note that in our case, the tree level contribution to the $\rho$ parameter is negligibly small since we consider a very small $v_L$ \cite{Czakon:1999ue}.}.  In the minimal LRSM, the new scalars enter the self energies and consequently, $\delta \rho$ depends on their spectrum  \cite{Czakon:1999ue}. In our analysis we use \textbf{SPheno} to calculate $\delta \rho$ and find that its typical size is less than $10^{-4}$, which lies within the experimentally allowed region $-0.000313 \leq \delta \rho \leq  0.00156$ \cite{Baak:2014ora,Olive:2016xmw}.

\item The vacuum stability condition, $\lambda_1 \geq 0$, $\rho_1 \geq 0$, $\rho_1 + \rho_2 \geq 0$ and $\rho_1 + 2 \rho_2 \geq 0$, which was derived in
  \cite{Chakrabortty:2013zja,Chakrabortty:2013mha}.

\item Unitary constraints on the quartic scalar couplings.  Since the theory predicts eleven complex fields and seventeen independent quartic couplings, the unitary constraints are rather complicated in a general case. In our analysis, however, we note that $\alpha_3$ and $\kappa_1$ are much larger than the other couplings. So we consider a simple case that only $\alpha_3$ and $\kappa_1$ among the couplings are nonzero. We work in the basis
    ($\chi \chi^\ast$, $H_L H_L$, $A_L A_L$, $H_L^\pm H_L^\mp$, $H_L^{\pm \pm} H_L^{\mp \mp}$, $H_R H_R$, $A_R A_R$, $H_R^\pm H_R^\mp$, $H_R^{\pm \pm} H_R^{\mp \mp}$, $H_1 H_1$, $A_1 A_1$, $H_1^\pm H_1^\mp$, $H_2 H_2$, $A_2 A_2$, $H_2^\pm H_2^\mp$), and calculate all $2 \to 2$ transition amplitudes as did in
    \cite{Lee:1977eg,Akeroyd:2000wc,Cao:2013wqa}. After the diagonalization of the transition matrix, we obtain the following unitary condition:
    \begin{eqnarray}
    4 \kappa_1^2 + 9 \alpha_3^2 + \sqrt{16 \kappa_1^4 + 28 \kappa_1^2 \alpha_3^2 + 15 \alpha_3^4} \leq (8 \pi)^2.
    \end{eqnarray}
    We note that the unitary constraint was also discussed in \cite{Chakrabortty:2016wkl}, where the authors sorted out all possible quartic contact terms in terms of the physical fields where the vertex factors of each coupling are linear functions of the quartic couplings. Finally, the authors required each of the couplings to be less than $8\pi$. Obviously, the limits obtained in this way are rather conservative, which can be seen from their results
    \begin{eqnarray}
    \alpha_1 \leq 8 \pi, \quad \alpha_2 \leq 4 \pi, \quad \alpha_1 + \alpha_3 \leq 8 \pi, \quad \cdots.
    \end{eqnarray}

\end{itemize}

\begin{table}
\begin{center}
\begin{tabular}{lr@{\hspace{5em}}lr}\toprule
$\kappa_1$ 		& 2.0 \		            & $\rho_1$	     & $1.4\times 10^{-2}$	\\
$\rho_2$  		& $1.1\times 10^{-2}$\	 	            & $\rho_3$	     & $7.3\times 10^{-2}$  \\
$m_{H_1^0}$ 		& $1.25\times 10^2 $\, {\rm GeV}	 	& $m_{H_2^0}$	     & $3.01 \times 10^{3}$\, {\rm GeV} \\
$m_{H_3^0}$	    & $3.34\times 10^3 $\, {\rm GeV}	    & $m_{H_4^0}$	     & $2.00 \times 10^{4}$\, {\rm GeV} \\
$m_{A_1}$ 	    & $3.01 \times 10^3 $\, {\rm GeV}  	    & $m_{A_2}$	     & $2.00 \times 10^{4}$\, {\rm GeV} \\
$m_{H_1^\pm}$   & $3.01 \times 10^3 $\, {\rm GeV}	    & $m_{H_2^\pm}$  & $2.00 \times 10^{4}$\, {\rm GeV} \\
$m_{H_1^{\pm\pm}}$& $2.99 \times 10^3 $\, {\rm GeV} 	    & $m_{H_2^{\pm\pm}}$&$3.01\times 10^{3}$\, {\rm GeV} \\
$m_{Z^{'}}$	    & $1.57\times 10^4 $\, {\rm GeV} 	    & $m_{W^{'}}$	 & $9.37\times 10^{3}$\, {\rm GeV} \\
$\Gamma_{Z^{'}}$& $4.66\times 10^{2}$\, {\rm GeV} 	    & $\Gamma_{W^{'}}$&$3.27 \times 10^{2}$\, {\rm GeV} \\
$Br(Z^{'}\rightarrow H_{1,2}^{++}H_{1,2}^{--})$  & $\sim2.8\%$	& $Br(W^{'+}\rightarrow H_{1}^{+} H_2^0)$	&$\sim5.1\times 10^{-5}$\\
$Br(H_1^{++}\rightarrow l_{i}^{+}l_{i}^{+})$ 	 & $\sim33{\%}$, $33{\%}$, $33{\%}$ & $Br(H_2^{++}\rightarrow l_{i}^{+}l_{i}^{+})$  &$\sim33{\%}$, $33{\%}$, $33{\%}$ \\
$Br(H_1^{+}\rightarrow l_{j}^{+}\nu_{i})$ 	 & $\sim33{\%}$, $33{\%}$, $33{\%}$	& $Br(H_2^0 \rightarrow \nu_{i} \nu_{i})$ & $\sim33{\%}$, $33{\%}$, $33{\%}$ \\
\bottomrule
\end{tabular}
\caption{A benchmark point in Scenario II. Here the nearly degenerated particles
$H_2^0$, $A_1$, $H_1^\pm$ and $H_1^{++}$ correspond to triplet scalars, and $H_4^0$, $A_2$ and $H_2^\pm$ are
Bi-doublet scalars. }
\label{Point-property}
\end{center}
\end{table}

\begin{figure}[t]
\centering
\includegraphics[height=5.5cm,width=7cm]{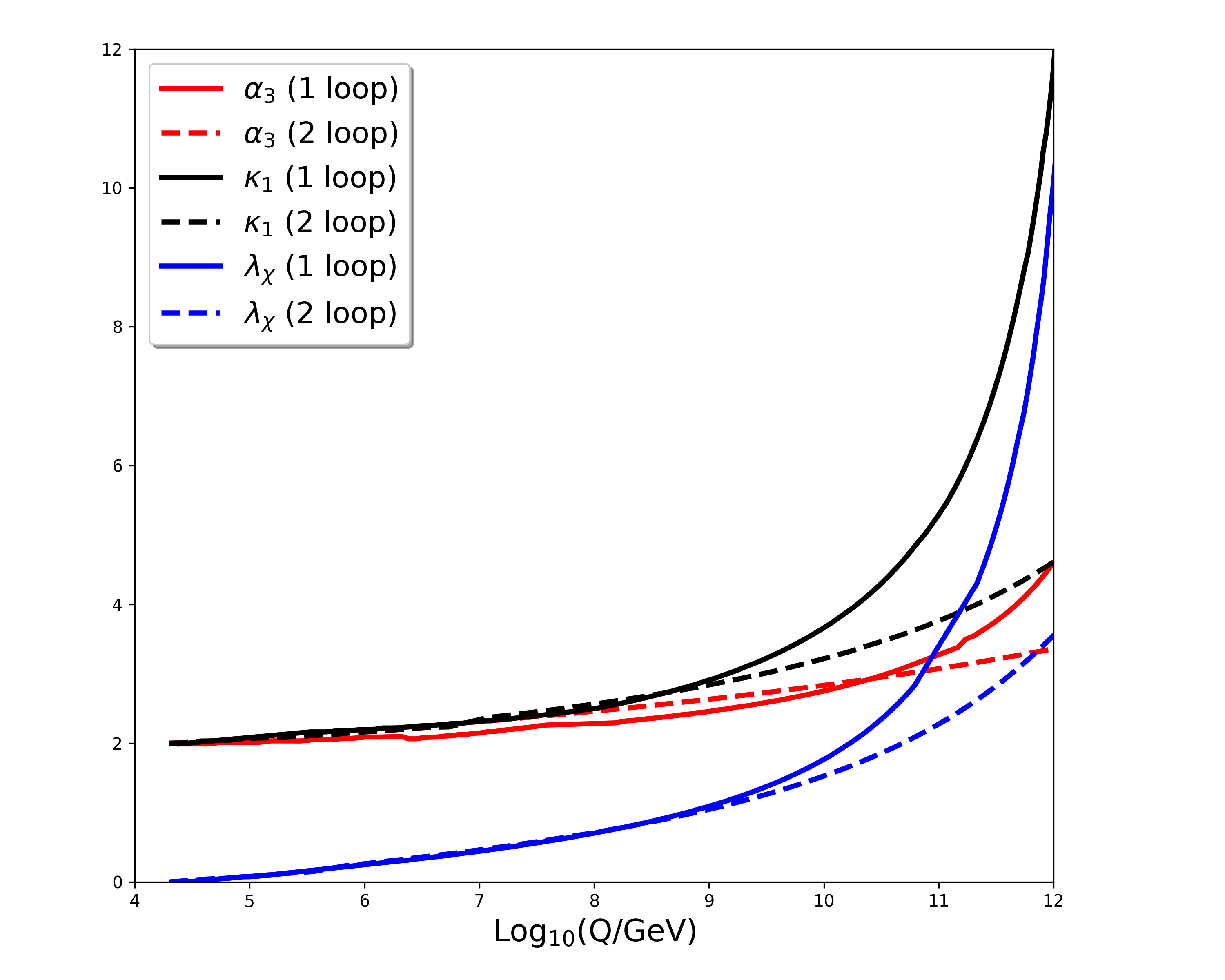} \hspace{-0.cm}
\includegraphics[height=5.5cm,width=7cm]{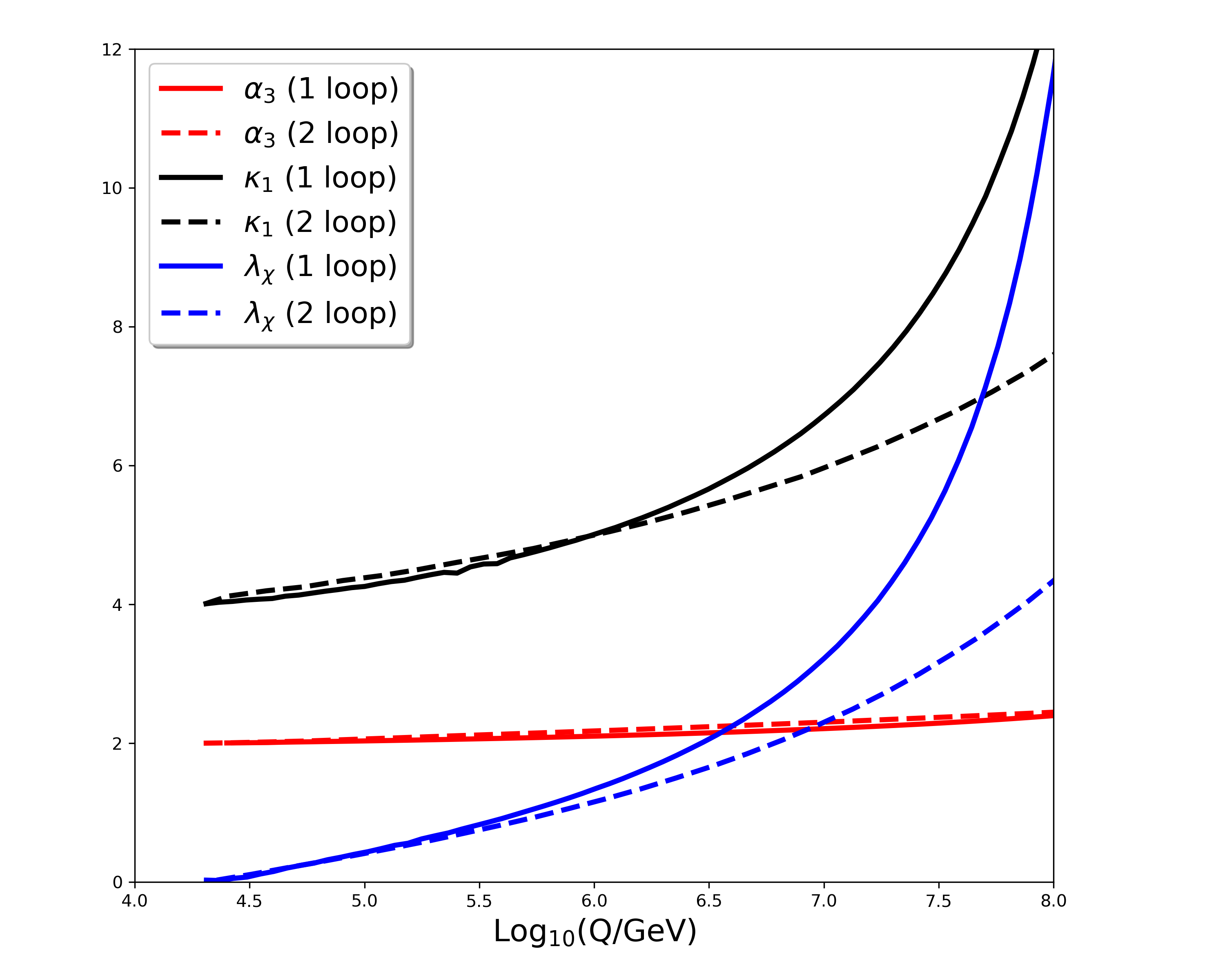}
\vspace{-0.4cm}
\caption{ \text{\bf Left} panel: the RGE evolution of the three largest couplings, $\kappa_1$ (black lines), $\lambda_\chi$ (blue lines) and $\alpha_3$ (red lines), for the benchmark point of Scenario II in Table \ref{Point-property}. In calculating the RGE, only the dominant contributions are included, and the solid and dashed lines correspond to one-loop and two-loop results, respectively. \text{\bf Right} panel: same as the left panel, but for $\kappa_1 = 4$ which is favored
by Scenario I to explain the DAMPE excess. \label{RGE} }
\end{figure}

\section{\label{Section-4}Implication of the explanation}

\begin{figure}[t]
\centering
\includegraphics[height=5.5cm]{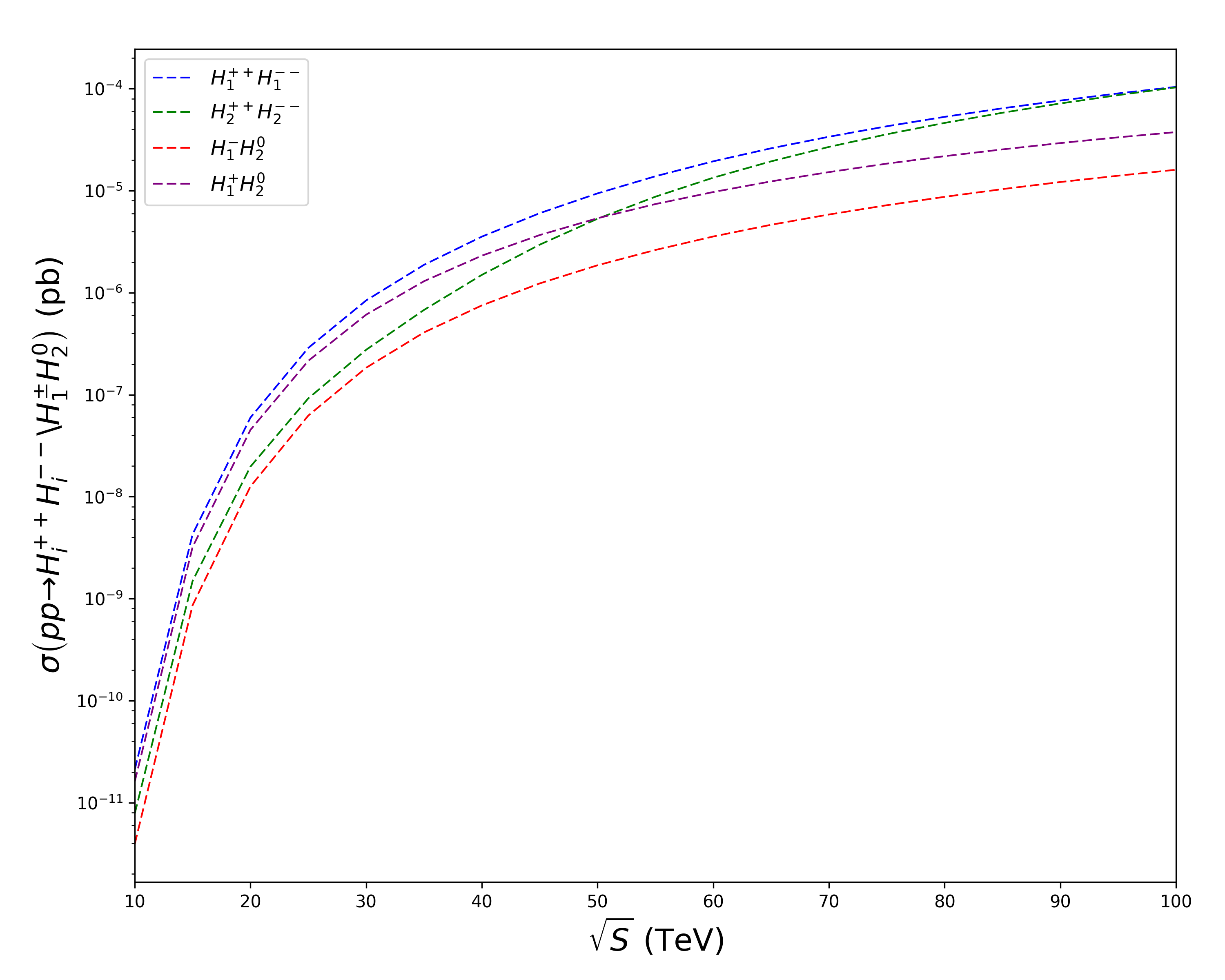}
\caption{Production rates of $\sigma(p p \to H_1^{++} H_1^{--}, H_2^{++} H_2^{--}, H_1^+ H_2^0, H_1^- H_2^0)$
for the benchmark point of Scenario II in Table \ref{Point-property} as a function of the collision energy $\sqrt{s}$.  \label{Collider}}
\end{figure}

From the discussion in Section \ref{Section-3}, one can learn that our explanation relies on a moderately large $\kappa_1$, i.e.
$\kappa_1 \sim 4$ for Scenario I and $\kappa_1 \sim 2$ for Scenario II. This may spoil the perturbativity of the theory.
We investigate this issue by first choosing one benchmark point in Scenario-II and presenting
its properties in Table \ref{Point-property}. Then we study the behaviors of some potentially large couplings with the increase
of energy scale according to the renormalization group equations (RGE) presented in the Appendix.
In the left panel of Fig.\ref{RGE}, we show the results of the benchmark point with the solid and dashed lines denoting one-loop and two-loop predictions,
respectively. This panel indicates that $\kappa_1$ as the
largest coupling is the first one to reach its Landau pole at the scale near $4 \times 10^{12} \,{\rm GeV}$ at one-loop level, while at two-loop level the increase rate of
$\kappa_1$ to higher energy scale is greatly reduced. In more details, for the energy scale lower than about $10^{9} {\rm GeV}$ where $\kappa_1 \simeq 3$, the difference induced by the one-loop and two-loop $\beta$ functions is negligibly small. However, for energy scale higher than about $10^{10} {\rm GeV}$ where $\kappa_1 \simeq 4$, the difference between one-loop and two-loop is significant. This fact
implies that the perturbativity of the theory becomes worsened greatly and perturbative calculations are no longer reliable near $10^{10}$ GeV.
As a comparison, we choose another $\kappa_1 =4$ (favored by Scenario I to explain the excess) while keeping $\alpha_3,\lambda_\chi$ the same, and study
its RGE behavior. We find that $\kappa_1$ reaches its Landau pole at the scale $3.2 \times 10^8 \,{\rm GeV}$ at one-loop level as shown in the right panel of Fig.\ref{RGE}. Similar to the case in the left panel, the sharp increasing behavior at one-loop level is ameliorated after including the two-loop corrections near $10^7$ GeV.
We should note that the appearance of Landau pole may not be problematic. Possible new particles, such as heavy fermions or new gauge bosons, which can contribute to the beta function at high energy (such as the Pati-Salam unification scale), may change the RGE behavior of the quartic couplings upon such energy scale. We also remind that the calculation of the two-loop results is rather involved, and we obtain our results by utilizing the package PyR@TE \cite{Lyonnet:2013dna,Lyonnet:2015jca,Lyonnet:2016xiz}.
 To the best of our knowledge, the two-loop result in the minimal LRSM is still absent in the literature.

The discussion in Section \ref{Section-3} also indicates that the explanation requires one or more doubly
charged Higgs $H_{1,2}^{\pm\pm}$ with masses around 3 TeV. These particles, once produced at future collides,
will decay dominantly into same sign lepton pairs with invariant mass peaking around 3 TeV. Given that such
a signal is quite distinct at hadron collides, we briefly discuss its observability in future experiment.
In Fig.\ref{Collider}, we show for the benchmark point in Table \ref{Point-property} the production rates $\sigma(pp \to H_1^{++} H_1^{--}, H_2^{++} H_2^{--})$ which proceed mainly through exchanging vector bosons
such as $\gamma,Z,Z^\prime$, as well as through neutral Higgs bosons $H_i$ but with
smaller contributions. This figure indicates that the production rates increase with the
collision energy $\sqrt{s}$, but are generally small even for $\sqrt{s} = 100 \,{\rm TeV}$ which has
$\sigma\sim 0.1 \,{\rm fb}$. This implies that even if one neglects the SM background arising from possibly mis-tagged leptons, the
integrated luminosity of $\gtrsim \mathcal{O}(100) \,{\rm fb}^{-1}$ may be needed to detect the signal.

In Fig.\ref{Collider} we also show the production rates $\sigma(p p \to H_1^+ H_2^0, H_1^- H_2^0)$.
Since $H_1^-$ and $H_2^0$ mainly decay into $\ell\nu$ and neutrino pairs respectively,
these processes will result in $\mathrm{mono-}\ell+E_{miss}^T$ signal which has the same signature as $WZ$ associated
production and thus is also distinct. Nevertheless, since the corresponding production rates are too small
(at most $10^{-2} \,{\rm fb}$), the prospect of detecting them may be dim.

For the benchmark point in Table \ref{Point-property}, we also checked that the production rates of $\sigma(p p \to H_1^{++} H_2^{--}, H_1^+ A_1^0)$ are much smaller than those of above production channels. Moreover, we note that the doubly charged Higgs $H_{1,2}^{\pm\pm}$ may contribute to the process $e^+ e^- \to \ell_i^+ \ell_j^-$ via $t$-channel mediation at future International Linear Collider (ILC). However,
the signal strength of this process is proportional to $Y_\Delta^4$. Due to the small value of
$Y_\Delta$ as well as the heavy mass of the $H_{1,2}^{\pm\pm}$, we estimate that the effect is negligible.

\section{\label{Section-5}Conclusion}

Given the fact that the electroweak interaction in the SM violates parity while all other interactions conserve parity, Left-Right symmetric model (LRSM) has been an attractive extension of the SM which can address the parity violation in EW interaction, generate tiny neutrino masses, accommodate DM candidates and provide a framework for baryogenesis through leptogenesis. In this work we enlarge the field content of the minimal LRSM by adding one gauge singlet scalar field, which acts as the simplest extension of the LRSM to include DM physics, and utilize the resulting theory to study the recently reported DAMPE results of cosmic $e^+e^-$ flux. We considered two scenarios to explain the DAMPE peak with a scalar DM $\chi$: 1) $\chi\chi^* \to H_1^{++}H_1^{--} \to \ell_i^+\ell_i^+\ell_j^-\ell_j^-$; 2) $\chi\chi^* \to H_{k}^{++}H_{k}^{--} \to \ell_i^+\ell_i^+\ell_j^-\ell_j^-$ accompanied by $\chi\chi^* \to H_1^+ H_1^- \to \ell_i^+ \nu_{\ell_i} \ell_j^- \nu_{\ell_j}$ with $\ell_{i,j}=e,\mu,\tau$ and $k=1,2$. We
fit the theoretical prediction on the $e^+e^-$ spectrum to relevant experimental data to determine the scalar mass spectrum favored by the DAMPE excess. We also consider
various constraints from theoretical principles, collider experiments as well as DM relic density and direct search experiments.
We find that there are ample parameter space which can interpret the DAMPE data while passing the constraints. Our interpretation, on the other hand, implies
the breakdown of the perturbativity of the theory at the energy scale ranging from $10^{7} \ {\rm GeV}$ to $10^{11} \ {\rm GeV}$, which can be avoided by the
intervention of other new physics.  We also discussed briefly collider signals of our explanation and concluded that an luminosity of $\gtrsim {\cal{O}} (100) \,{\rm fb^{-1}}$
is needed to probe the signal at future hadronic collider with $\sqrt{s} = 100 \,{\rm TeV}$.

Before we end this work, we would like to clarify two important points.

One is that the gauge singlet scalar DM considered in this work is motivated by some UV completions of the LRSM
and also by simplicity. The essential two ingredients of our interpretation of the excess include the existence of one or more leptophilic mediators $S_i$ which are
$H_{1,2}^{\pm \pm}$ and $H_1^\pm$ in the LRSM, and sufficiently large quartic scalar couplings $\chi \chi^\ast S_i S_i^\ast$ which ensure that the DM $\chi$ acquires right relic density without any suppression of the $\langle \sigma v \rangle_0$ for the annihilation $\chi \chi^\ast \to S_i S_i^\ast$. It should be noted that scalar leptophilic DM model containing $\phi_L,\phi_R$ introduced in the beginning of Section \ref{Section-3} also possesses the aforementioned ingredients, where the $\lambda_{\Delta_1}$ term in Eq.(64) and the $\lambda_{\phi \Delta \Delta \phi}$ term in Eq.(67) of \cite{Garcia-Cely:2015quu} play the same role as the $\kappa_1$ term in DM physics. Therefore in principle the model may also be utilized to explain the excess albeit that its structure is much more complex. Obviously, if the DM explanation of the excess is confirmed by future data, a careful examination of the model should be carried out, which is beyond the scope of this work.

Another point is the physical enrichments of this work compared to our earlier studies \cite{Cao:2017ydw,Cao:2017sju}, where we extended the SM by lepton specific symmetries and built
anomaly free theories in an economic way. In all of these works we consider a scalar DM candidate to generate the $e^+e^-$ flux by a two step annihilation process, followed by solving the propagation equation of electron/positron in cosmic ray and seeking for the maximum value of the likelihood function to attain the $\chi^2$ map on $\Delta m - m_\chi$ plane needed to produce the right shape of the $e^+ e^-$ spectrum. Then we obtain the parameter space of the model relevant to explaining the excess by considering the constraints from DM relic density and its direct detection experiments. Due to their shared workflow procedures, we organize the discussions in a similar way. However, the underlying physics in these works are quite different, which is reflected in the following aspects:
\begin{itemize}
\item The theoretical frameworks considered in \cite{Cao:2017ydw,Cao:2017sju} focus on the DAMPE excess, and the DM candidate together with the lepton-specific gauge bosons are added by hand. By contrast, in this work we consider the physically well-motivated LRSM where the leptophilic mediators and the DM candidates can arise quite naturally without the needs for lepton-specific gauge interactions. This model, aside from being capable of explaining the DAMPE excess, can also account for many fundamental problems in particle physics such as neutrino masses and baryogenesis. Moreover, as we mentioned in the Introduction, the theory has rich phenomenology at colliders which will be test in future. Consequently, it is of particular interest.

\item In earlier works \cite{Cao:2017ydw,Cao:2017sju} we extend the SM by certain lepton specific gauge symmetries in an economic way. In doing this, we note that the anomaly free condition puts non-trivial requirement on the quantum number of leptons for the new symmetry and consequently, the new gauge boson $Z^\prime$ as the mediator of the two step annihilation must decay in certain pattern, e.g. either democratically into three generations of lepton pairs or into $e^+e^-$ and $\mu^+\mu^-$ with equal branching ratios. By contrast, in this work we consider the triplet scalars as the mediators of the DM annihilation. The decay modes of the scalars are determined by the Yukawa coupling $Y_\Delta$ which depends on the parameters $v_{L,R}$ and $M_D$ and is therefore somewhat arbitrary. Moreover, since the LRSM predicts six triplet Higgs bosons, we have more choices in selecting the mediators of the annihilation than those in \cite{Cao:2017ydw,Cao:2017sju}. These features make the discussions of this work more adaptive to future cosmic ray data.

\end{itemize}

\section*{Appendix}

Here we present the RGEs of the parameters in the extended minimal LRSM discussed in this work. Since the complete forms of the equations in the LRSM are quite
complicated \cite{Chakrabortty:2016wkl,Rothstein:1990qx} \footnote{One can also get the total one-loop $\beta$ functions of the minimal LRSM from the website https://github.com/jlgluza/LR.},
we consider the case of large $\kappa_1$, $\lambda_\chi$ and $\alpha_3$ and only include potentially large contributions in
the RGEs. At the initial stage of this work, we calculated the effects of $\kappa_1$ and $\lambda_\chi$ on the one-loop $\beta$ functions of the couplings by hand, and
took the rest contributions from \cite{Rothstein:1990qx}. While revising the manuscript, we noticed that the package PyR@TE \cite{Lyonnet:2013dna,Lyonnet:2015jca,Lyonnet:2016xiz} can calculate the $\beta$ functions automatically. Thus we implement the model into the package to calculate all the one-loop $\beta$ functions. We find that, as far as the dominant contributions listed below are concerned, the two sets of results agree with each other. We also calculate the two-loop $\beta$ functions for the couplings $\kappa_1$, $\lambda_\chi$, $\alpha_3$ and $\alpha_1$ using the package and keep only the dominant terms. We note that the calculation of the two-loop contributions is rather computationally heavy, and attaining the complete set of the two-loop results is beyond the capability of our computing resources.

The resulting RGEs are given by
\begin{eqnarray}
  16\pi^2\frac{\partial g_s}{\partial t} &=& -7 g_s^3, \quad 16\pi^2\frac{\partial g_2}{\partial t} = -\frac{7}{3}g_2^3, \quad 16\pi^2\frac{\partial g_{BL}}{\partial t} = \frac{14}{3} g_{BL}^3\\
    16\pi^2\frac{\partial y_t}{\partial t} &=& y_t (-\frac{2}{3} g_{BL}^2-\frac{9}{2} g_2^2-8g_s^2+8y_t^2)\\
  16\pi^2\frac{\partial \kappa_1}{\partial t} &=& 4 \kappa_1^2+ 8\lambda_{\chi}\kappa_1 +16 \rho_1\kappa_1+6\rho_3 \kappa_1-12\kappa_1 g_2^2-6\kappa_1 g_{BL}^2\\ \nonumber
  &&+ \bf{\frac{1}{16\pi^2}(-4 \alpha_1^2 \kappa_1 - 3 \alpha_3^2 \kappa_1 - 15 \kappa_1^3 -
 48 \kappa_1^2 \lambda_{\chi} - 40 \kappa_1 \lambda_{\chi}^2 - 4 \alpha_1 \alpha_3 \kappa_1)} \\
  16\pi^2\frac{\partial \lambda_{\chi}}{\partial t} &=& 20 \lambda_{\chi}^2+6\kappa_1^2 + \bf{\frac{1}{16\pi^2}(-24 \kappa_1^3 - 60 \kappa_1^2 \lambda_{\chi} - 240 \lambda_{\chi}^3)} \\
  16\pi^2\frac{\partial \lambda_1}{\partial t} &=& 32 \lambda_1^2+ \frac{5}{2} \alpha_3^2 + 16 \lambda_1\lambda_3 +16 \lambda_3^2+6\alpha_1\alpha_3 + 6\alpha_1^2  \\ \nonumber
   && + 12 \lambda_1 y_t^2-6 y_t^4 -18\lambda_1 g_{2}^2+ 3 g_2^4\\
  16\pi^2\frac{\partial \lambda_3}{\partial t} &=& - \alpha_3^2+ 24 \lambda_1 \lambda_3  + 16\lambda_3^2 +12 \lambda_3 y_t^2+3 y_t^4 - 18\lambda_3 g_2^2 + \frac{3}{2} g_{2}^2\\
  16\pi^2\frac{\partial \alpha_1}{\partial t} &=& 8 \lambda_1\alpha_3+ \alpha_3^2 +8 \lambda_3\alpha_3+16\rho_1 \alpha_3+8\rho_2\alpha_3+3\rho_3\alpha_3+20\lambda_1 \alpha_1\\ \nonumber
  &&+8\lambda_3\alpha_1 +16\rho_1\alpha_1 +8\rho_2\alpha_1+6\rho_3\alpha_1+4\alpha_1^2+6\alpha_1 y_t^2\\ \nonumber
  &&-6\alpha_1 g_{BL}^2-21\alpha_1 g_2^2 +6g_2^4 \\ \nonumber
  &&+ \bf{\frac{1}{16\pi^2}(-18 \alpha_1^3 - 10 \alpha_1^2 \alpha_3 - \alpha_1 \kappa_1^2 -
 6 \alpha_3^3 - \frac{27}{2} \alpha_1 \alpha_3^2)} \\
  16\pi^2\frac{\partial \alpha_3}{\partial t} &=& 4 \lambda_1\alpha_3+ 4\alpha_3^2 -8 \lambda_3\alpha_3+4\rho_1 \alpha_3-8\rho_2\alpha_3+8\alpha_1\alpha_3+6\alpha_3 y_t^2\\ \nonumber
  && -6\alpha_3 g_{BL}^2-21\alpha_3 g_2^2 \\ \nonumber
  &&+\bf{\frac{1}{16\pi^2}(-34 \alpha_1^2 \alpha_3 - 34 \alpha_1 \alpha_3^2 -
  \frac{19}{2} \alpha_3^3 - \alpha_3 \kappa_1^2)} \\
  16\pi^2\frac{\partial \rho_1}{\partial t} &=& 2 \alpha_3^2+ 28 \rho_1^2 + 16\rho_1 \rho_2  + 16\rho_2^2 +3\rho_3^2 +4\alpha_1 \alpha_3 +4\alpha_1^2    \\ \nonumber
  && -12\rho_1 g_{BL}^2+6g_{BL}^4 +12 g_{BL}^2 g_2^2 -24 \rho_1 g_2^2 + 9 g_2^4 + \kappa_{1}^2\\
  16\pi^2\frac{\partial \rho_2}{\partial t} &=& - \alpha_3^2+ 24 \rho_1\rho_2 + 12\rho_2^2 -12 \rho_2 g_{BL}^2 - 24\rho_2 g_2^2 -12 g_2^2 g_{BL}^2 + 3g_{2}^4 \\
   16\pi^2\frac{\partial \rho_3}{\partial t} &=& 2 \alpha_3^2+32\rho_1 \rho_3+16\rho_2 \rho_3+4\rho_3^2+8\alpha_1 \alpha_3+8\alpha_1^2-12\rho_3 g_{BL}^2  \\ \nonumber
   &&-24\rho_3 g_2^2 +12g_{BL}^4+2\kappa_1^2
\end{eqnarray}
where $y_t \simeq (Y_{Q_1})_{33} $ in the small $\tan \beta$ limit denotes top quark Yukawa coupling, $t = \ln \mu$, and the terms in bold denote dominant two-loop contributions.
Note that since the operators ${\rm Tr} (\Phi^{\dag} \Phi) \left[ {\rm Tr} (\Delta_L \Delta_L^{\dag}) +
{\rm Tr} (\Delta_R \Delta_R^{\dag})\right ] $ and $\left[ {\rm Tr}(\Phi \Phi^{\dag} \Delta_L \Delta_L^{\dag}) + {\rm
Tr}(\Phi^{\dag} \Phi \Delta_R \Delta_R^{\dag}) \right] $ contain the same fields, the calculation of the $\beta$ functions for $\alpha_1$ and that for $\alpha_3$ are entangled.
Thus we also present the two-loop result of $\alpha_1$, although it is not important in our numerical calculation.
From these analytic expressions, one can learn that for the benchmark point in Table \ref{Point-property}, the two-loop contribution is at most $6\%$ of its corresponding one-loop result at TeV scale, although the values of $\kappa_1$ and $\alpha_3$ are quite large.

\section*{Acknowledgement}

We thank J. Chakrabortty from Indian Institute of Technology for the usage of the package PyR@TE.
This work is supported by the National Natural Science Foundation of China (NNSFC) under grant No. 11575053 and 11675147, and also
by the Innovation Talent project of Henan Province under grant number 15HASTIT017, by the Young Core instructor of the Henan education department.


\bibliographystyle{JHEP}

\end{document}